# Towards Frontier Safety Policies 'Plus'

**Matteo Pistillo**\*†
Apollo Research

**Executive Summary**

This paper examines the state of affairs on Frontier Safety Policies ('FSPs') in light of capability progress and growing expectations held by government actors and AI safety researchers from these company-led safety policies. It subsequently argues that FSPs should evolve to a more fine-grained version, which this paper calls 'FSPs Plus.' Compared to the first wave of FSPs led by a subset of frontier AI companies (Anthropic's Responsible Scaling Policy, OpenAI's Preparedness Framework, and Google DeepMind's Frontier Safety Framework), FSPs Plus should be built around two main pillars. First, FSPs Plus should adopt a new, reasonably comprehensive, well-defined set of metrics—precursory capabilities. In this respect, this paper recommends that international (e.g., ISO, IEC) or domestic (e.g., NIST, CEN-CENELEC, BIS) standardization bodies develop a standardized taxonomy of precursory components to high-impact capabilities that FSPs Plus could then adopt by reference. While these bodies engage different industry, government, and academia stakeholders in their quest towards standardization, the Frontier Model Forum could lead the way by establishing preliminary consensus amongst frontier AI developers on this topic. Second, FSPs Plus should expressly incorporate AI safety cases and establish a mutual feedback mechanism between FSPs Plus and AI safety cases. To establish such a mutual feedback mechanism, FSPs Plus could be updated to include a clear commitment to make AI safety cases at different milestones in AI systems' development and deployment, to build and adopt safety measures based on the content and confidence of AI safety cases, and—also on this basis—to keep updating and adjusting FSPs Plus.

\* Correspondence to matteo@apolloresearch.ai.
† I am grateful to Charlotte Stix, Mikita Balesni, Saad Siddiqui, Joe O'Brien, Alejandro Ortega, and the Safe AI Forum ('SAIF') Red Lines Working Group participants for thoughts and comments. All comments were made in personal capacity and do not represent endorsement of the views expressed in this paper. Mistakes are my own.

Pre-print. Under review.

# I.     Introduction

This paper focuses on Frontier Safety Policies ('FSPs') and how they can and should be improved. FSPs are voluntary risk management policies developed and adopted by a subset of frontier AI developers ([Anthropic, 2024](#); [OpenAI, 2023](#); [Google DeepMind, 2024](#)) with the goal of keeping the risks deriving from frontier AI development and deployment within acceptable levels ([METR, 2023](#); [METR, 2024](#)). The core idea behind FSPs is evaluation-gated scaling ([Hubinger, 2023](#)). These policies have two main components. First, companies *pre-define* the conditions under which it could be too dangerous to continue developing or deploying AI systems without additional safety measures ([METR, 2023](#); [Karnofsky, 2024(a)](#); [Karnofsky, 2024(b)](#)). These conditions are identified through levels of capabilities—also referred to as 'capability thresholds' ([Koessler et al., 2024](#); [Anthropic, 2024](#)) or 'tripwire capabilities' ([Karnofsky, 2024(a)](#)). In other words, FSPs use capability levels as a proxy for risk—absent protective measures, certain capability levels are presumed to pose unacceptable or intolerable risks ([Frontier Model Forum, 2024](#)). Second, if the predefined capability thresholds are reached, companies *commit* to eliminating or reducing the relevant potential risks by applying adequate risk mitigation measures or—should the protective measures appear insufficient— stalling further scaling and deployment ([Hubinger, 2023](#)).

Industry, regulators, and AI safety researchers have used various terminologies to describe these company-led scaling policies, including 'preparedness frameworks' ([OpenAI, 2023](#)), 'safety frameworks' ([Google DeepMind, 2024](#); [Kasirzadeh, 2024](#)), and 'safety and security frameworks' ([EU General-purpose AI Code of Practice](#)). In this paper, I use the term 'Frontier Safety Policies' or 'FSPs' to refer to all these different examples of company-led risk management frameworks ([METR, 2024](#)).

This paper considers criticism leveled at existing FSPs by the research community and argues that FSPs should now evolve into FSPs Plus. FSPs Plus should be built around two main pillars: (1) FSPs Plus substitute capability levels with a standardized taxonomy of precursory capabilities; and (2) FSPs Plus incorporate AI safety cases expressly and establish a mutual feedback mechanism between FSPs and AI safety cases.

## II. Impact Statement

Research attention on FSPs is timely. Not only are FSPs likely here to stay as voluntary commitments ([Titus, 2024](#)), but they might also be folded into regulation ([Hubinger, 2023](#); [Alaga et al., 2024](#)). For instance, the latest draft of the [EU General-purpose AI Code of Practice](#) under Article 56 of the AI Act (second draft) includes a commitment that signatories "adopt[], implement[], and mak[e] available to the AI Office a Safety and Security Framework" ([Commitment 4](#)). Such Framework must contain "the risk management policies that the Signatory adheres to in order to assess and mitigate systemic risks from their general-purpose AI models" ([Commitment 4](#))—essentially, the core elements of FSPs. Further, during the 2024 Seoul Summit, 16 companies committed to developing governance frameworks ([Commitment IV](#)) that would help them assess the risks posed by AI models and systems ([Commitment I](#)), define thresholds at which risks are unacceptable unless adequately mitigated ([Commitment II](#)), and identify and apply risk mitigations ([Frontier AI Safety Commitments, 2024](#))—in other words, governance frameworks that would contain the core elements of FSPs. Under the Frontier AI Safety Commitments, the deadline to publish such FSP-like safety frameworks is the [AI Action Summit](#), which will take place in Paris, France, on February 10-11, 2025. 17 Chinese AI developers have also recently taken a similar initiative through a set of voluntary AI Safety Commitments ([CAICT, 2024](#)). These companies have committed to "build[ing] safety and security risk management mechanisms," including by "defin[ing] realistic safety risk baselines" and implement[ing] risk management practices throughout the entire AI development and deployment lifecycle" ([Commitment I](#)).

Working on stronger FSPs is also warranted by capability progress. We have witnessed very significant capability progress within a short time period, the pace of which is only expected to increase. This considerable jump in capabilities is outpacing existing FSPs. For instance, when OpenAI published its [Preparedness Framework](#) in December 2023, the latest model publicly available was GPT-4. Not only do the company's subsequent models—including o1 and the announced o3—meaningfully surpass GPT-4 in capabilities ([OpenAI o1 System Card](#); [ARC Prize, 2024](#)) and hence pose greater risk ([Guan et al., 2024](#)), but they also highlight dangerous capabilities that are not encompassed in the Preparedness Framework, such as scheming ([Apollo Research, 2024](#); [OpenAI o1 System Card](#); [Google DeepMind, 2024](#)). At the same time, some sets of capabilities are getting easier and cheaper to generate ([Karnofsky, 2024(a)](#)), enlarging the number of 'frontier AI developers.' For instance, DeepSeek recently launched DeepSeek-V3—a frontier model that purportedly surpasses GPT-4o and Claude 3.5 Sonnet on several benchmarks at the aggregated training cost of only USD 5.576M ([DeepSeek-V3 Technical Report](#))—and DeepSeek-R1, which purportedly shows performance comparable to OpenAI's o1 ([DeepSeek-R1](#)).

## III. Criticisms Towards Existing FSPs

This Section reviews the main criticisms leveled against FSPs. Section IV will advance two policy solutions to address these criticisms. Since Anthropic first introduced its Responsible Scaling Policy in 2023 ([Anthropic, 2023](#)), these policies have been both commended and criticized. For example, experts have given low grades to existing FSPs ([FLI AI Safety Index 2024](#); [Alaga et al., 2024](#)), observed that they deviate significantly from risk management norms ([Campos et al., 2024](#)), and suggested that FSPs be substantially revised ([Kasirzadeh, 2024](#)). More specifically, the criticisms leveled at FSPs can be grouped into three main categories, discussed in Sections III.A-C below.

A.      FSPs Should Be More Specific

The first category of criticism concerns the FSPs' level of detail. Two aspects of FSPs have been considered insufficiently specific. First, the capability levels identified in existing FSPs are defined only loosely (Kasirzadeh, 2024) and lack sufficient clarity, which could lead to different interpretations of the capability thresholds (Anderson-Samways et al., 2024) and ultimately prevent them from functioning as effective triggers for risk mitigation strategies (Campos et al., 2024). For instance, observers could reasonably differ on what 'meaningfully improved assistance' means in the context of Chemical, Biological, Nuclear, and Radiological ('CBRN') capabilities in OpenAI's Preparedness Framework (Titus, 2024). Further, researchers argue that capability thresholds in FSPs should not contain undefined upper bounds or be open-ended (Kasirzadeh, 2024). For instance, Anthropic is still "working on defining any further Capability Thresholds that would mandate ASL-4 Required Safeguards" (Anthropic, 2024)—in other words, its highest-level capability thresholds are still undefined. Second, FSPs do not include clear mechanisms for regular updates (Titus, 2024).

B.      FSPs Should Be Broader

The second category of criticism concerns FSPs' scope. In short, the claim is that FSPs should be broader and contain more detailed risk identification. First, FSPs should contain more risk *levels* vertically (Anderson-Samways et al., 2024)—including today's risks (Titus, 2024)—and more risk *types* horizontally (Anderson-Samways et al., 2024). Delphi studies and Fishbone Diagrams could help identify uncovered risk types (Campos et al., 2024).

Second, the scope of FSPs should be widened to include elements of risk identification that are not currently part of the picture. It has been suggested that FSPs contain causal pathways to catastrophic risk (Kasirzadeh, 2024). Similarly, it has been recommended that FSPs consider deployment contexts and "clarify scenarios" (SaferAI, 2024). The latest draft of the EU General-purpose AI Code of Practice appears to echo these suggestions. Under the Code of Practice, signatories commit to "identify the pathways (series of interconnected events, conditions, or factors that lead to the manifestation of risks) by which the development and deployment of their general-purpose AI model with systemic risk could produce the systemic risks identified" (Measure 8.1). Signatories also commit to describe and justify "risk tiers" for each systemic risk through, for example, "harmful scenarios" (Measure 4.2). In a similar fashion, the Frontier AI Safety Commitments specify in a footnote that thresholds can be "defined using … deployment contexts." Finally, FSPs could include information on the probability of risks materializing. The EU General-purpose AI Code of Practice follows this approach. According to the latest draft, Signatories commit to including best-effort estimates of risk timelines in their frameworks, such as through "ranges or probability distributions over different scenarios and timelines" (Measure 4.4).

C.      FSPs Should Be More Verifiable

The third category of criticism of FSPs concerns their verifiability from the outside. First, FSPs are not perceived as sufficiently specific to be verifiable (Anderson-Samways et al., 2024; Section III.A above). A solution that has been advanced is to set precise quantitative risk thresholds (Kasirzadeh, 2024; *see*

Koessler et al., 2024) and define absolute risk above a given baseline (Anderson-Samways et al., 2024; Campos et al., 2024). This broadly aligns with recent recommendations from policymakers, including the Department for Science, Innovation and Technology's recommendation to define and operationalize risk thresholds through "specific, testable observations," "such that multiple observers with access to the same information would agree on whether a given threshold had been met" (Emerging Processes for Frontier AI Safety).

Second, FSPs are not perceived as sufficiently harmonized across the industry to be cross-verified. While some frontier AI companies have adopted structured—albeit still improvable—FSPs, others have not committed to any precautionary measures (FLI AI Safety Index 2024). Further, substantial disparities exist amongst existing FSPs, rendering cross-verification challenging. In short, the field lacks a harmonized approach to frontier safety policies. The latest draft of the EU General-purpose AI Code of Practice echoes this need. Under the Code of Practice, signatories commit to identifying risk tiers at which—absent appropriate mitigations—the level of risk would be unacceptable and, "where possible," "clearly define[]" risk tiers "on a fixed and comparable scale across Signatories" including by "align[ing] with best practice and international approaches" (Measure 4.2).

## IV. Evolution of FSPs to FSPs Plus

Section III reviewed and summarized the three main criticisms that have been leveled against FSPs and how these frontier safety policies have failed to meet growing expectations. These criticisms concern FSPs' level of detail (Section III.A), scope (Section III.B), and external verifiability (Section III.C). Regarding the level of detail, researchers criticized that FSPs are not sufficiently clear and detailed, are susceptible to disparate interpretations, and contain undefined upper bounds. Researchers also observed that FSPs do not contain clear mechanisms for regular updates. Regarding FSPs' scope, researchers suggested that FSPs include more risk levels and consider causal pathways to catastrophic risk, probability, and deployment scenarios. Regarding the verifiability of FSPs, researchers recommended that FSPs be more specific and harmonized across the industry to be verifiable.

This Section IV advances recommendations to meet the expectations described under Section III and increase FSPs' robustness in light of capability advancement. In particular, this paper argues that FSPs should evolve into FSPs Plus, built around two pillars: a standardized taxonomy of precursory capabilities (Section IV.A) and an explicit feedback mechanism with AI safety cases (Section IV.B). These two Sections follow an analogous structure. First, I describe each pillar. Second, I explain why the pillar could meaningfully advance FSPs. Third, I examine what activities are necessary to achieve each pillar. Finally, I describe reasonable counter-arguments that have been raised or could be raised against each pillar.

### A. FSPs Plus Adopt a New Set of Standardized Metrics: Precursory Capabilities

FSPs Plus should abandon divergent and under-specified capability thresholds and adopt a new, reasonably comprehensive set of standardized metrics: precursory capabilities.

**(1)** *The First Pillar*

The AI safety research community should develop a taxonomy of precursory components to high-impact capabilities. These precursory capabilities should be standardized through a consensus-driven process led by international or domestic standardization bodies. The taxonomy should be as comprehensive as reasonably possible at the time it is drafted. It should then be regularly updated as our understanding of AI capabilities improves.

Precursory capabilities are smaller preliminary components to high-impact capabilities that an AI model needs to have in order to unlock more advanced capabilities (Pistillo & Stix, 2024). They are 'but for' skills—skills without which a certain action by an AI model is simply impossible (Pistillo & Stix, 2024). In this sense, they are causally connected to one another in a spectrum that leads from basic core capabilities to high-impact capabilities and, hence, from 'less close' to 'closer' to possible catastrophic risk (Pistillo & Stix, 2024). Since they fragment a high-impact capability into smaller components, precursory capabilities can serve as penultimate redlines. Figure 1 below provides an illustrative sketch of some precursory capabilities to AI scheming. The blue boxes describe illustrative examples of precursory capabilities to AI scheming, and the red boxes explain why these precursory capabilities are causally connected.

Figure 1 — Illustrative Example of Precursory Capabilities and Their Causal Connection

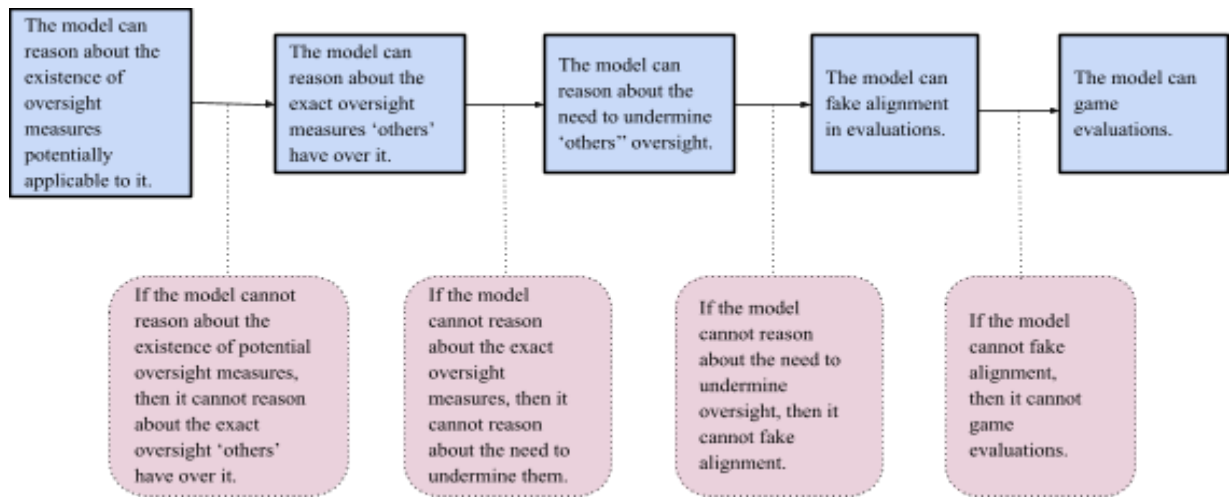

**(2)** *Supporting Arguments*

Reaching international or national standards on a taxonomy of precursory capabilities could be an optimal solution to mitigate existing FSPs' lack of specificity, insufficient depth of vertical risk levels, and external verifiability (Sections III.A-C above).

A taxonomy of precursory capabilities has advantages over under-specified capability thresholds. First, precursory capabilities provide a clearer metric to test for and build consensus than current capability thresholds. Second, a precursory capabilities taxonomy could provide a more granular and exhaustive perspective on risk levels and future capabilities, as each high-impact capability would be broken down into increasingly dangerous precursory components. In this sense, it could also serve as a best-guess forecasting tool. Third, less ambiguity in the conditions triggering safety measures would make FSPs more easily verifiable by external observers.

Achieving a global or national standard on the described taxonomy also has some advantages. First, not only would the core metric of FSPs be clearer and more specific, but it would also be harmonized across AI companies and, possibly, across jurisdictions, enabling international recognition ([NIST AI 100-5](#)). Second, a standardization process like the ones followed by the International Organization for Standardization ('ISO') or the National Institute of Standards and Technology ('NIST') would ensure that the taxonomy represents the consensus reached by experts from various stakeholders, including industry, academia, and government ([ISO, Developing Standards](#); [NIST AI 100-5](#)).

Furthermore, a standardization process would also align with various calls from government actors and industry. For instance, in the United States, President Trump's [Executive Order 13859](#) of February 11, 2019, on Maintaining American Leadership in Artificial Intelligence, stated that "[t]he United States must drive the development of appropriate technical standards and reduce barriers to the safe testing and deployment of AI technologies," and listed the development of "international standards" as a "strategic goal" ([Sec. 2](#)). Acting on this Executive Order, NIST prepared its Plan for Federal Engagement in Developing Technical Standards and Related Tools, in which it also recommended the development and adoption of "metrics to quantifiably measure and characterize AI technologies" ([NIST, 2019](#)). President Biden's [Executive Order 14110](#) of October 23, 2023, on the Safe, Secure, and Trustworthy Development and Use of Artificial Intelligence—recently revoked by President Trump ([The White House, 2025](#))—also called for the "development and implementation of AI-related consensus standards" ([Section 11(b)](#)). NIST then adopted a plan calling for "outreach to and engagement with international stakeholders and standards-developing organizations to help drive the development and implementation of AI-related consensus standards" ([NIST AI 100-5](#)). Within the matters "[u]rgently needed and ready for standardization," NIST included "[m]easurement methods and metrics," such as "[s]hared testing, evaluation, verification, and validation (TEVV) practices for AI models and systems" ([NIST AI 100-5](#)). A taxonomy of precursory capabilities could fall within these metrics. Similarly, in the European Union, the Commission has already argued for the importance of standard-setting efforts in AI and entrusted this task to the European Committee for Standardization ('CEN') and the European Committee for Electrotechnical Standardization ('CENELEC') ([Commission Implementing Decision, 2023](#); [Annex 1, 2023](#)). In industry, OpenAI very recently published its Economic Blueprint for 2025 ([OpenAI Economic Blueprint, 2025](#)). In this document, the company expressed "support" for the "development of standards and safeguards, ensur[ing] that such standards are recognized and respected by other nations and international bodies on behalf of the US private sector," and observed that "developers and users have a responsibility to follow *clear, common-sense standards* that keep the AI roads safe" ([OpenAI Economic Blueprint, 2025](#), emphasis added).

**(3)**    *Next Steps*

Two steps separate us from reaching this first pillar. The first step is reaching a standardized taxonomy of precursory capabilities. The ideal final output is an official "standard" ([ISO/IEC Guide 2:2004](#)) that details precursory components to each high-impact capability on a spectrum from less close to closer to high-impact capabilities. FSPs and future legal frameworks could then incorporate this standard by reference. This first step can be broken down into two sub-tasks. First, standard drafters could identify which capabilities qualify as 'high-impact capabilities' and the precursory components necessary to reach them. Efforts in this direction have already been made by, among others, NIST with its AI Risk Management Framework ([NIST-AI-600-1](#)) and the EU General-purpose AI Code of Practice ([Measure 3.2](#)). Second, standard drafters could reach a consensus on the precursory components of the identified high-impact capabilities. A previous paper has proposed an initial, illustrative sketch of precursory components of scheming ([Pistillo & Stix, 2024](#)). That sketch could be expanded and refined for each high-impact capability.

Other scholars previously noted that, within a couple of years, there could be a growing emphasis on formal industry standards ([Karnofsky, 2024(a)](#)). This process could have two stages. In the first stage, industry members could build consensus on precursory capabilities through the Frontier Model Forum, whose core objectives include enabling standardized evaluations of capabilities ([Frontier Model Forum](#)). This first stage could be started today. In the second stage, the standardization process could be continued and perfected through national or, preferably, international standardization bodies. At the national level, this task could be undertaken, for instance, by NIST in the United States, by CEN and CENELEC—and their relevant technical committees, such as CEN/CLC/JTC 21—in the European Union, and by the British Standards Institution ('BSI') in the United Kingdom. At the international level, this process could be undertaken by global standard development organizations such as ISO and the International Electrotechnical Commission ('IEC') through their relevant committees and subcommittees focusing on AI, such as ISO/IEC JTC 1/SC 42. As NIST observed, international standards should be preferred to avoid friction between potentially incompatible domestic standards ([NIST, 2019](#)). Besides, domestic standardization bodies like NIST already participate in developing relevant international standards ([NIST, AI Standards](#)), and global standards like ISO or IEC are usually adopted and recognized domestically ([Garrido et al., 2024](#)).

The second step is updating FSPs based on the standardized precursory capabilities (or drafting FSPs based on this new set of metrics for companies that have not yet adopted them). In line with the two elements of FSPs (described in Section I above), AI companies could: (i) identify the precursory capabilities that, absent protective measures, are presumed to pose unacceptable or intolerable risks; and (ii) pre-commit to adequate protective measures if such precursory capabilities are reached.

**(4)** *Opposing Arguments*

The main argument against FSPs' first pillar is that a standardized taxonomy of precursory capabilities is not feasible or worth the AI research community's effort, considering our shared understanding of capabilities and the pace of capabilities progress. In other words, a taxonomy of precursory capabilities is impossible or impractical to achieve because our understanding of capabilities is limited, and capabilities scale and emerge too fast. As a result, a precursory capabilities taxonomy would be inaccurate and incomplete, and a joint effort by industry, government, and academia would not be able to reach a better specification of capabilities than the ones proposed by some frontier AI companies ([Anthropic, 2024](#); [OpenAI, 2023](#); [Google DeepMind, 2024](#)). Experts might also reasonably differ on which precursory capabilities lead to catastrophic risk and never reach a consensus or not reach it in meaningful time. A taxonomy would also require frequent updating to reflect advancements in capabilities and our understanding of AI systems. Standardization would only exacerbate these concerns because it typically takes a long time, often years. For example, developing an ISO standard usually takes about three years ([ISO, Developing standards](#)). Given the above, specifying existing capability thresholds (and even standardizing them) may be easier and more effective than devising a novel taxonomy. For instance, companies could reach a consensus on a more detailed definition of 'meaningfully improved assistance' rather than breaking down CBRN capabilities into their causally connected precursory components (OpenAI's [Preparedness Framework](#); *see* Section III.A above).

Indeed, identifying causally-connected precursory capabilities is not an easy task. Nonetheless, industry, government, and academia could start by focusing on what they agree on rather than what they disagree on and use their initial—even partial—consensus as the basis for upgrading FSPs. In other words, experts could still reach a consensus on areas within the taxonomy with a higher degree of certainty. Reflecting on precursory capabilities could also be a great exercise for defining better threat models. Expanding on the standardization consideration, while it is true that standardization processes like ISO/IEC could take considerable time, AI companies could also reach consensus through more 'agile' avenues, such as the Frontier Model Forum. Then, domestic and international standardization bodies could build off this initial consensus, involving additional stakeholders and implementing and perfecting the taxonomy. An initial taxonomy could also be developed through 'leaner' consortia and subsequently incorporated into an ISO/IEC standard. This happened, for instance, with the SERT standard developed by the Standard Performance Evaluation Corporation ([SPEC, SERT Suite](#)).

### B. FSPs Plus Should Incorporate AI Safety Cases and Establish a Mutual Feedback Mechanism

This Section describes the second pillar of FSPs Plus. FSPs Plus should expressly incorporate AI safety cases and establish a mutual feedback mechanism between FSPs and AI safety cases.

In this paper, 'AI safety case' refers to a new version of safety cases devised for frontier AI systems. AI safety cases are structured, evidence-based rationales that an AI system deployed to a specific setting is unlikely to cause catastrophic outcomes ([Clymer et al., 2024](#); [Buhl et al., 2024](#); [Balesni et al., 2024](#); [Goemans et al., 2024](#)). In other words, AI safety cases aim to demonstrate through evidence that the likelihood of catastrophic outcomes from an AI system's given application in a given environment is below an acceptable threshold ([Defence Standard 00-56 Part 1](#); [Clymer et al., 2024](#); [Irving, 2024](#); [Buhl et al., forthcoming](#)). Leading scientists have supported the development of safety cases before model deployment ([IDAIS Venice](#)), and the UK AI Safety Institute ('AISI') is currently working on safety cases ([UK AISI](#)). However, frontier AI companies are yet to publish their first AI safety cases.

**(1)** *The Second Pillar*

FSPs already rely on AI safety cases implicitly, at least in relation to one of the four arguments generally advanced in AI safety cases—inability arguments ([Goemans et al., 2024](#); [Buhl et al., 2024](#); [Grosse, 2024](#)). In other words, FSPs rely on an implicit argument that AI systems are safe because they are not capable of accomplishing the tasks described by FSPs' capability thresholds ([Buhl et al., forthcoming](#)). The second pillar of FSPs Plus is making this 'silent' connection between FSPs and AI safety cases more explicit and robust.

The following is a sketch of the commitments that FSPs Plus could include to this end. First, FSPs Plus should include a clear commitment to run AI safety cases—and hence demonstrate through evidence that systems are safe—for each new AI system at different milestones ([Buhl et al., forthcoming](#)). This paper does not dig deep into which milestones should trigger the commitment to run a safety case. Intuitively, a safety check could occur, at least, before deciding on further development, before deploying a model internally, and before deploying it externally. This commitment should also encompass the post-deployment phase by committing to showing that an AI safety case continues to hold if operated in an out-of-scope environment ([IDAIS Venice](#)).

Second, AI companies should clearly commit to regularly updating FSPs Plus based on the content and confidence of AI safety cases. In other words, FSP safety measures should be fortified if an AI company cannot persuasively demonstrate that an AI system's risk level is sufficiently low (for instance, if the confidence is too low). AI safety cases could highlight that an FSP's safeguard measures are, in fact, insufficient to make a system safe in a specific scenario. For instance, a safety case might illustrate that, with capability progress, the ASL-3 Security Standard that is part of Anthropic's Responsible Scaling Policy for autonomous AI R&D does not adequately mitigate the risk ([Anthropic, 2024](#); *see* [Hobbhahn, 2025](#) on short timelines). Of course, this example is purely illustrative but highlights how AI safety cases could shed light on grey areas where FSPs' safeguards are insufficient and improve preparedness ([Clymer et al., 2024](#)).

Third, FSPs Plus should include a clear commitment to consider AI safety cases and take action based on the arguments' robustness and confidence. Ideally, this commitment would go one step further than soft guidelines specifying how safety cases are assessed ([Clymer et al., 2024](#)). FSPs Plus could explicitly commit to not scale or deploy further until an AI safety case proves the system is safe ([Buhl et al., 2024](#)). Coupled with the first pillar of FSPs Plus (*see* Section IV.A above), this second commitment would entail pausing development or deployment if certain standardized precursory capabilities are found and AI safety cases cannot make a solid inability or control argument for why the system is safe ([Buhl et al., 2024](#)). Illustrative drafting of an FSPs Plus could be: "If we find that one of our AI systems is plausibly capable of [*a standardized precursory capability*], at a time when we are not in a position to [*demonstrate through an AI safety case that the system is safe*], then we will take the following measures to contain the risk …" (adapted from the wording suggested by [METR, 2024](#)).

**(2)** *Supporting Arguments*

Assuming that AI safety cases move from safety sketches to higher-confidence guarantees (*see* Section IV.B(3) below), incorporating AI safety cases and establishing a clear feedback mechanism could improve FSPs and mitigate some criticisms against them regarding their scope and verifiability (*see* Sections III.B and III.C above).

First, the second pillar of FSPs Plus could remedy the lack of specificity and external verifiability ([Anderson-Samways et al., 2024](#)), especially if coupled with FSP Plus' first pillar (Section IV.A above).

FSPs Plus would refer to a standardized taxonomy of precursory capabilities and contain a commitment to take certain safety measures if some of these precursory capabilities are reached. AI safety cases would then explain why the standardized precursory capabilities either do not exist in the examined system or are under control in the specific scenario. In other words, in their FSPs Plus, AI companies would adopt a standardized set of metrics (precursory capabilities) that they commit to test their systems for; then, in AI safety cases, these companies would develop legible safety arguments based on the evaluation of these standardized metrics.

Second, the second pillar of FSPs Plus could address the recommendation that FSPs consider deployment contexts (*see* [SaferAI](); [Measure 4.2, EU General-purpose AI Code of Practice](); [Frontier AI Safety Commitments, 2024]()) and identify the pathways to catastrophic outcomes (*see* [Kasirzadeh, 2024](); [Measure 8.1, EU General-purpose AI Code of Practice]()). When preparing an AI safety case, companies should consider realistic settings and justify their development or deployment decisions ([Clymer et al., 2024]()). This includes justifying AI companies' claims about an AI system's safety in a specific deployment setting—including users, online services, and physical infrastructure—that the system interacts with ([Clymer et al., 2024](); [Balesni et al., 2024]()).

Third, the second pillar of FSPs Plus could remedy the absence of a risk probability assessment in FSPs (*see* [Measure 4.4, EU General-purpose AI Code of Practice]()) since AI safety cases aim to describe the probability of unacceptable outcomes ([Balesni et al., 2024]()).

Finally, the second pillar of FSPs Plus could solve the lack of a mechanism for regular updates in FSPs ([Titus, 2024]()) and move past a generic reference to 'periodical' updates ([Google DeepMind, 2024]()). Establishing a clear commitment to running safety cases at certain milestones in a system's development and deployment and updating FSPs Plus based on the content of AI safety cases could establish a clear timeline for FSPs Plus updates.

**(3)** *Next Steps*

The second pillar of FSPs Plus cannot happen overnight. Two significant intermediary steps are necessary. First, advancements must be made in state-of-the-art AI safety cases. While the AI safety community has called for the development of "high-confidence" safety cases ([IDAIS Venice]()), when compared to other sectors, such as nuclear or aviation, current AI safety cases are still 'best-guess' safety cases, operating more as tools for reasoning transparency than as high-confidence quantified guarantees ([Hobbhahn, 2024]()). In other words, it is not currently possible to guarantee an AI system's real-world safety ([Grosse, 2024]()) and provide a reliable numerical quantification of confidence ([Barrett et al., forthcoming]()), to the point that some members of the research community argue that even the term 'safety cases' is inappropriate. AI safety cases are also time-bound, as it is impossible to build safety cases for AI models that are much more advanced than current ones ([Irving, 2024]()). Second, AI companies should upgrade their FSPs to FSPs Plus by incorporating and further specifying the commitments described under previous Section IV.B(1).

**(4)** *Opposing Arguments*

A reasonable argument against FSPs Plus' second pillar is that AI safety cases may not be able to provide sufficient confidence in the safety of AI systems within a reasonable time frame and, therefore, may not be adopted by frontier AI developers as hoped. For instance, researchers have argued that frontier AI companies have less than a 20% chance of succeeding at making high-assurance safety cases and that they are unlikely to pause scaling if they cannot guarantee such high confidence ([Greenblatt, 2025](#)). This shortcoming could bottleneck development and deployment decisions that depend on AI safety cases in FSPs Plus or result in frontier AI developers simply not respecting their FSP commitments tied to AI safety cases. In other words, one could align with the growing section of the AI research community that already considers AI safety cases a 'failed experiment' and argue that FSPs should not incorporate AI safety cases as this instrument is not likely to be successful in the future.

To address this issue, FSPs could be updated to incorporate AI safety cases only if and once frontier AI developers publish their first AI safety cases. While sketches have been made, AI companies are yet to publish their first comprehensive AI safety case. Besides, while it is possible that AI safety cases will not instill full confidence in the safety of AI systems within a short timeline, legible safety arguments outlining the result of relevant evaluations would still help justify and further improve FSPs. In other words, this potential limitation should not affect the first two commitments described in Section IV.B(3) above. The rollout of the third recommended commitment—i.e., considering AI safety cases and taking development and deployment actions based on their outcome—could be more gradual to accommodate the state of the art of AI safety cases.

If, by contrast, AI safety cases do succeed in providing high-confidence safety guarantees, then one could also argue that FSPs would no longer be needed. In other words, if AI safety cases provided mathematical evidence of AI systems' safety, FSPs Plus would no longer serve a clear purpose. This argument feels inconsistent with the purpose of FSPs and AI safety cases. AI safety cases complement and operationalize FSPs rather than replace them. For example, AI safety cases cannot perform one of the main functions of FSPs: forecasting future capabilities and pre-committing to certain protective actions once certain capabilities are reached. In other words, FSPs are intended to be forward-looking, while AI safety cases are not. In this sense, FSPs Plus would serve as a necessary 'glue' or 'mortar' between AI safety cases.

## V. Conclusions

In conclusion, the time is right for FSPs to start evolving into FSPs Plus. Recent initiatives such as the Frontier AI Safety Commitments and the EU General-Purpose AI Code of Practice suggest that FSPs are here to stay—at least on a voluntary basis, if not in regulatory form (Sections I-II). After considering criticism leveled at FSPs regarding their level of detail (Section III.A), scope (Section III.B), and verifiability (Section III.C), this paper recommended building FSPs along two main pillars.

First, **FSPs Plus should adopt a uniform taxonomy of precursory capabilities as more granular triggers for AI companies' safety commitments** (Section IV.A). This taxonomy should be as comprehensive as reasonably possible, list all precursory components to high-impact capabilities in a spectrum from less close to closer to possible catastrophic risks, and be standardized. The standardization process should develop in two parallel steps. AI companies should start working towards a consensus on precursory capabilities through 'agile' avenues, such as the Frontier Model Forum. In the meantime, international and/or domestic standardization bodies—such as ISO, IEC, NIST, CEN-CENELEC, and BSI—should gather experts from different stakeholders (industry, government, academia) and work towards a more refined text. The end goal is to achieve a standardized breakdown of high-impact capabilities into causally connected precursory components that AI companies could then incorporate by reference into their FSPs Plus and rely on in their AI safety cases.

Second, **FSPs Plus should incorporate AI safety cases and establish clear feedback mechanisms between FSPs Plus and AI safety cases** (Section IV.B)—which means including in FSPs Plus a commitment to run AI safety cases, consider AI safety cases and take action based on their outcome, and keep updating FSPs Plus based on AI safety case content and confidence. To fully operationalize this pillar, further research is essential to improve AI safety cases from best guesses to high-confidence safety guarantees.